Cluster trajectory of SOFA score in predicting mortality in sepsis

Ke Yuhe1, Matilda Swee Sun Tang2, Celestine Jia Ling Loh3, Hairil Rizal Abdullah1,3, Nicholas Brian Shannon4

1 Division of Anaesthesiology and Perioperative Medicine, Singapore General Hospital, Singapore

2 Royal Adelaide Hospital, Adelaide, Australia

3 DukeNUS Medical School, Singapore

4 Division of General Surgery, Singapore General Hospital, Singapore

Corresponding Author: Ke Yuhe

Email: yuhe.ke@mohh.com.sg

Address: Singapore General Hospital, Division of Anaesthesiology and Preoperative Medicine. Outram Rd, Singapore 169608.

Declaration of interest: There are no conflicts of interest.

Abstract

"Failure to rescue" is the failure in recognizing and responding to patient experiencing a potentially preventable complication, and it underscores the critical challenge of averting patient mortality. Swiftly identifying high-risk individuals within the intensive care unit (ICU) is pivotal to enhancing patient outcomes. Yet, the performance of prevailing risk prediction scores remains limited, yielding overly pessimistic prognostications for high-risk patients who could potentially respond favorably to interventions, and concurrently offering overly optimistic evaluations for low-risk patients who might unexpectedly decline. The widely employed Sequential Organ Failure Assessment (SOFA) Score, aimed at tracking ICU patient progress, holds promise for surmounting these limitations by incorporating serial SOFA score measurements.

To bridge this research gap, our study is dedicated to conceiving and validating an innovative risk prediction model. This model's core objective is to refine the precision of identifying patients susceptible to the specter of "failure to rescue" within the ICU.

Methods

We analysed 3,253 patients from the MIMIC IV database. Criteria included adherence to sepsis-3 guidelines, admission via the emergency department, a minimum of 72 hours of ICU residency, and active resuscitation status. Leveraging advanced techniques including group-based trajectory modeling coupled with dynamic time warping and k-means clustering, we unveiled four distinctive trajectory archetypes within dynamic SOFA scores.

Results

These trajectory profiles encompassed Cluster A, characterized by consistently low scores; Cluster B, demonstrating a rapid ascent followed by decline; Cluster C, illustrating higher baseline scores evolving gradually toward amelioration; and Cluster D, displaying sustained elevation. Cluster D exhibited the lengthiest stays in both ICU and hospital settings, accompanied by the highest ICU and hospital mortality rates. Cluster A and B shared similar ICU discharge rates, while Cluster C displayed initially comparable rates yet exhibited a more gradual transition to ward settings.

Conclusion

Consistent monitoring of dynamic shifts in SOFA scores emerges as a valuable asset for evaluating sepsis severity and gauging therapy response, potentially facilitating timely interventions and elevating overall patient care quality.

Introduction:

Sepsis, a life-threatening condition resulting from a dysregulated host response to infection, poses a significant burden on healthcare systems worldwide [1]. It is associated with high morbidity and mortality rates, making it one of the leading causes of in-hospital deaths. In intensive care units (ICUs), sepsis-related mortality rates range from 10% to 40% [2]. Given the severity and prevalence of sepsis, accurately predicting patient outcomes is crucial for optimising care delivery and resource allocation.

The Sequential Organ Failure Assessment (SOFA) score [3] has been established as the diagnostic criterion for sepsis (Sepsis 3.0) [4] and serves as an effective tool to assess the severity of organ dysfunction and predict mortality in the ICU setting. However, the SOFA score is static and fails to account for dynamic changes in a patient's response to initial resuscitation. Previous studies have indicated that monitoring the difference in SOFA scores at two-time points can predict the prognosis of sepsis patients at 28 days [5]. Despite this insight, little is known about the trajectory of the SOFA score over time and its relationship with clinical outcomes. This knowledge gap highlights the need for a comprehensive understanding of SOFA score trends during the critical initial 72 hours of ICU admission and their association with mortality rates and LOS.

To address this research gap, big data analytics and advanced machine learning algorithms offer a promising avenue for improving sepsis care. Previous studies have successfully applied clustering techniques to identify trajectory patterns in blood pressure measurements in emergency departments [6]. By employing similar clustering methods to identify trajectory patterns within SOFA scores, it becomes possible to identify subgroups of patients with similar trends and explore the differences in clinical outcomes among these groups. The knowledge gained from such analyses has the potential to guide clinical decision-making and ultimately enhance patient outcomes.

Therefore, the objective of this study is to investigate the association between dynamic changes in SOFA scores over the first 72 hours of ICU admission and sepsis patient mortality rates and LOS. By harnessing the power of big data analytics and machine learning techniques, we aim to gain valuable insights into the progression of organ dysfunction and the likelihood of adverse outcomes. These insights will contribute to the development of more effective interventions and personalised treatment strategies, ultimately leading to improved patient care and outcomes in sepsis management.

Methods:

*Data source and selection of participants*

The dataset for this study is the Medical Information Mart for Intensive Care (MIMIC-IV) database [7] and is an updated version of MIMIC-III. This is a large, freely available database of de-identified electronic health records for patients admitted to ICU at the Beth Israel Deaconess Medical Center. The dataset includes demographic information, vital signs, laboratory values, medications, procedures, diagnosis, clinical outcomes and amongst other variables.

For our study, we extracted data on patients according to the diagnostic criteria of Sepsis 3. Patients are defined as having sepsis if they have 1) suspected infection combined, and 2) an acute increase in the SOFA score ≥ 2. We included patients who were admitted directly from the emergency department (ED) with ICU stay > 72 hours. Patients were excluded if they are younger than 18 years, had a do not resuscitate order or were transferred from another ICU, did not stay in ICU for more than 72 hours or had at least 3 consecutive missing SOFA score records in the first 72h after admission.

*Variables and endpoints*

The following baseline demographics of patients with sepsis were extracted: age, sex, race, marital status, language, insurance status, discharge location, LOS, use of mechanical ventilation and Charlson Comorbidity Index (CCI). The values were expressed in median and interquartile range (IQR) and number (%). The first set of blood test results on admission to ICU were extracted. The values were expressed in mean and standard deviations. Outcomes including mortality during the hospital stay, mortality within ICU, length of stay (LOS) in the hospital and ICU, and readmission during the hospital stay, were collected. The discharge time from ICU to the wards was noted and cut-offs at 7-day and 14-day were taken.

Time series data of SOFA scores were collected for each unique ICU admission. We then employed group-based trajectory modelling with dynamic time warping and k-means clustering to identify distinct trajectory patterns in dynamic SOFA scores.

*Group-based Trajectory Modelling (GBTM) with dynamic time warping (DTW)*

GBTM with DTW) was employed to analyse the trajectories of SOFA scores in this study. DTW, an algorithm widely used for time series analysis, was utilised to measure the similarity between two-time series that may vary in length and speed. By applying DTW to the SOFA score trajectories, we aimed to align them in a meaningful way, allowing for the identification of common features and patterns. The identified trajectory groups were analysed and interpreted based on the patterns observed in the SOFA score trajectories. Each group was described in terms of its initial SOFA score, peak score, response to treatment, and overall trajectory pattern.

*Clustering*

We employed a multi-step clustering approach to analyse the trajectory patterns of SOFA scores in sepsis patients. The methodology consisted of three key components: k-means clustering [8], selective clustering by cluster entropy, and locally weighted evidence accumulation (LWEA) as a consensus function [9]. To initiate the clustering process, we utilised the k-means algorithm, which is a commonly used partitioning-based clustering technique. The k-means algorithm partitioned the SOFA score trajectories into k distinct clusters, aiming to group similar trajectories together. Cluster entropy measures the degree of uncertainty within a cluster based on the distribution of trajectories. We selectively merged or split clusters based on their entropy values, ensuring that the resulting clusters were well-defined and internally homogeneous. LWEA is a version of hierarchical agglomerative clustering that combines information from multiple clustering ensemble members. It assigns weights to each ensemble member based on their performance in representing the underlying structure of the data. The clusters from different ensemble members were merged iteratively, considering their weighted evidence to create a consensus clustering solution.

Descriptive statistics were calculated based on the patient characteristics in the different clusters. Continuous variables were expressed as median and IQR (25-75th percentile) and their p-values evaluated using Kruskal-Wallis One-Way ANOVA test. Categorical variables were expressed as numbers and proportions and evaluated using Chi-square test. Missing values were imputed using the median for continuous variables and mode for categorical ones.

Univariable and multivariable linear regression models were performed to investigate the association between the different cluster groups and outcomes. The effect size was reported as an odds ratio (OR) and its 95% confidence interval (CI). Multivariable regression models were done by adjusting for these confounders. Bonferroni correction was used to adjust for p-value in multivariable regression models.

Data was accessed from MIMIC-IV 2.1 via Google BigQuery, further analysis was carried out in python 3.10, trajectory modelling was carried out using the tslearn (v0.5.3.2) package, lifelines (v0.27.7) package was used for survival curves. The code for extraction of time series SOFA score and feature extractions used for the study can be found publicly available on Github at https://github.com/nbshannon/mimic-iv/tree/main/sofatrajectory.

Results:

The search identified 76,541 adult ICU admissions from the MIMIC-IV database. A total of 34,790 fulfilled the sepsis-3 criteria. After exclusions, a total of 3,253 patients were included in the study (Figure 1).

The trajectory analysis of the time series SOFA score data revealed the presence of distinct patterns within the patient population, leading to the identification of four distinct clusters (Figure 2). The clusters were identified as Cluster A (n=990), Cluster B (n=387), Cluster C (n=1265) and Cluster D (n=611).

*Cluster characteristics*

Cluster A corresponds to patients who initially presented with a low SOFA score and maintained a consistently low level throughout their ICU stay. The peak SOFA score in this group reached a maximum of 3 points, indicating relatively mild organ dysfunction. This group likely experienced less severe infections and responded well to the initial treatment.

Cluster B represents patients who exhibited a rapid increase in their SOFA score within the first 24 hours of ICU admission. However, these individuals responded favourably to the administered treatments, leading to a subsequent decline in their SOFA score. The quick reduction in organ dysfunction suggests an effective response to therapy, potentially indicating a less severe infection or an early intervention.

Cluster C represents a cohort of patients who started with a higher baseline SOFA score, reaching a peak of 9 within the first 24 hours. Despite the severity of organ dysfunction, these patients still responded to the treatment provided, albeit at a slower rate. Over the next 2-3 days, their SOFA score gradually decreased, indicating a gradual improvement in their condition. This group likely experienced more severe infections but demonstrated a positive response to therapeutic interventions.

Cluster D comprises non-responders to treatment, characterised by persistently elevated SOFA scores. These patients did not exhibit significant improvements in their organ dysfunction throughout their ICU stay, suggesting a lack of response to the administered therapies. This group represents a subset of patients who experienced a more severe infection and may require alternative treatment strategies or further investigation to identify the underlying causes of their non-response.

The overall trend and distribution of SOFA scores were visually depicted in Figure 3. The plot revealed a notable decreasing trend in the scores after 24 hours of ICU admission. It is, however, important to acknowledge that the time 0 SOFA scores might introduce some inconsistencies inherent to the dataset (discussed in the limitations section).

*Baseline Characteristics*

The baseline characteristics of patients stratified based on the four trajectory groups are shown in Table 1. Similar to other studies[10], the higher SOFA scores in Clusters C and D can be correlated with the higher proportion of males in these clusters. There are no differences in the baseline ethnicity (p=0.904), and Cluster A had a statistically significant lower body mass index compared to the other clusters (p<0.05). Clusters A and B also had a significantly lower CCI score, white blood cells, creatinine, and bilirubin than clusters C and D (p<0.001). Platelets and bicarbonate were significantly higher in clusters A and B as compared to C and D (p<0.001). Taking all of these into account, this indicates that end-organ dysfunction was significantly milder in clusters A and B as compared to C and D.

*Outcomes*

Cluster D exhibited the most prolonged LOS in both the ICU and the hospital, along with a higher incidence of ICU deaths. In comparison, the LOS in the ICU and hospital was comparable among Clusters A, B, and C. However, Cluster C demonstrated a higher hospital mortality rate and a greater number of deaths within the ICU. Notably, while Clusters A and B displayed similar rates of discharge from the ICU, patients belonging to Cluster B had a higher incidence of ICU readmission within 30 days, as indicated in Table 2.

The days required for patients to transition from the ICU to the ward were graphically represented in Figure 4. Cluster D exhibited the lowest rates of discharge to the ward across all time points, indicating a prolonged stay in the ICU for these patients. Conversely, Clusters A and B displayed similar discharge rates from the ICU. Cluster C demonstrated discharge rates that were relatively comparable to those of Clusters A and B initially. However, after a two-week period of stay, the discharge rates for Cluster C patients noticeably decreased. This reduction suggests a slower transition of patients from Cluster C to the ward compared to the earlier time points.

Discussion:

This study aimed to investigate the trajectory patterns of patients with sepsis based on their time series SOFA score trajectories within the first 72 hours of ICU admissions. The analysis revealed four distinct clusters within the patient population, each representing a different trajectory of organ dysfunction and recovery.

Sepsis continues to be the leading cause of death among critically ill patients [4]. In the context of sepsis, monitoring the trajectory of the SOFA score dynamically can serve as a valuable tool not only for assessing the severity of sepsis but also for tracking responses to therapy. For our study, we established strict inclusion criteria focusing on patients with a DNR order who were directly admitted from the ED, rather than being transferred from another ICU beforehand. In another similar study, Yang et al. [10] conducted a comprehensive investigation using the MIMIC dataset and included all patients who met the sepsis-3 criteria. In their study, the trajectory groups were categorised into five groups, all of which demonstrated a plateauing SOFA score after 48 hours. This finding is likely attributable to the heterogeneity of the population described in their study. In contrast, our study revealed consistent improvements in clinical status during the initial 24-72 hours of ICU admission. This observation suggests that aggressive initial treatment of sepsis patients admitted from the emergency department to the ICU is crucial and can significantly enhance patient outcomes [11].

*Comparison of Clusters A and B*

Cluster A consisted of patients who maintained a consistently low SOFA score throughout their stay in the ICU while patients in Cluster B exhibited a rapid deterioration within the first 24 hours, which likely prompted intensivists to consider more aggressive therapeutic interventions, and responded well to them. The two groups of patients had similar ages and comorbidities with a mean CCI score of 5, but Cluster A had more females than Cluster B (46.6 vs 40.6%). These may point to a potential difference in how different genders respond to sepsis. Cluster B patients responded positively to the intensified treatments and were discharged from ICU faster although they have a higher peak of SOFA score. However, these patients also exhibited a higher frequency of readmissions. Although the overall mortality rates were similar between Cluster A and B, it is important to acknowledge that the utilisation of healthcare resources likely differed between the two clusters, a factor not addressed in this study.

*Comparison of Cluster C with Clusters A and B*

The patients in Cluster C started with a higher baseline SOFA score and reached a peak of 9 within the first 24 hours. They still responded to treatment, albeit at a slower rate, with a gradual decrease in their SOFA score over the next 2-3 days. This group experienced more severe infections[12] but demonstrated a positive response to therapeutic interventions. Cluster C patients were more likely to have mortality within ICU and hospital stay possibly contributed by the increased comorbidities and age[13]. This group of patients were also likely to present with worse septic markers and acidosis on admission to ICU.

*Cluster D*

Cluster D represented non-responders to treatment, as their SOFA scores remained persistently elevated, indicating persistent organ dysfunctions [12]. The SOFA score peaks at 10 within the first 24 hours of ICU admission, and the rate of falls is slow and gradual. The identification of this group of patients could either be used to guide prognosis and counselling, as well as early attempts to try more aggressive or alternative treatments which may reduce mortality [14].

One highlight from the result is that Clusters B, C and D had a higher proportion of males. This is consistent with previous research which postulated a difference in how different genders respond to inflammation [15]. The male sex hormones have been shown to be suppressive on cell-mediated immune responses while female sex hormones exhibit protective effects which may contribute to the natural advantages of females under septic conditions.

The clustering of patients into these distinct trajectory groups provides valuable information for understanding the heterogeneity of severity and responses to sepsis. These findings can guide clinical decision-making, treatment strategies, and resource allocation, ultimately leading to a more personalised and improved patient care. Similar methods of trajectory analysis rather than static measurements can also be applied to other inflammatory biomarkers [16] and vital signs such as shock index [17].

Limitations:

Several potential limitations and sources of bias should be considered due to the observational nature of this study. Firstly, the use of retrospective data from a single centre may limit the external validity of the findings, as it may not fully represent the diverse population of sepsis patients in other healthcare settings. Therefore, caution should be exercised when extrapolating these results to broader populations.

Another limitation stems from the exclusion of patients with less than 72 hours of ICU stay, as this may introduce bias into the results. By excluding patients who had shorter stays, the extremes of patient conditions—both those who were severely ill and those who recovered rapidly—are not captured within the sample population. Consequently, the findings may not fully reflect the outcomes of these patient groups, potentially affecting the generalisability of the results.

Furthermore, it is important to acknowledge the limitations associated with the time 0 SOFA scores. Inconsistencies in these scores can arise due to various factors, such as variations in the timing of SOFA score measurement relative to the onset of sepsis or potential delays in the documentation of initial scores. These limitations may introduce inaccuracies and affect the reliability of the time 0 scores. While these limitations and sources of bias are inherent to the study design, they should not undermine the significance of the results as the charting is reflective of real-world limitations.

Conclusions:

In conclusion, this study revealed distinct trajectory patterns of patients with sepsis based on their time series SOFA score trajectories within the first 72 hours of ICU admissions. The identification of four clusters provided valuable insights into the heterogeneity of organ dysfunction and recovery in sepsis patients. The findings emphasised the importance of monitoring the trajectory of the SOFA score dynamically as a tool for assessing the severity of sepsis and tracking responses to therapy.

Ethics Statement:

The MIMIC-IV database has been approved by the Massachusetts Institute of Technology and Beth Israel Deaconess Medical Center, and consent for the collection of the original data was provided by the patients (https://physionet.org/content/mimiciv/view-license/1.0/). Therefore, ethical approval and need for informed consent were waived for studies using data from this database.

Data Availability:

The MIMIC-IV data are available at https://mimic-iv.mit.edu/.

Funding:

This research did not receive any grant from funding agencies in the public, commercial, or not-for-profit sectors.

Conflicts of Interest:

No known competing interests.

Acknowledgements:

We wish to thank the Massachusetts Institute of Technology and Beth Israel Deaconess Medical Center for the MIMIC project.

Appendices:

Figure legend:

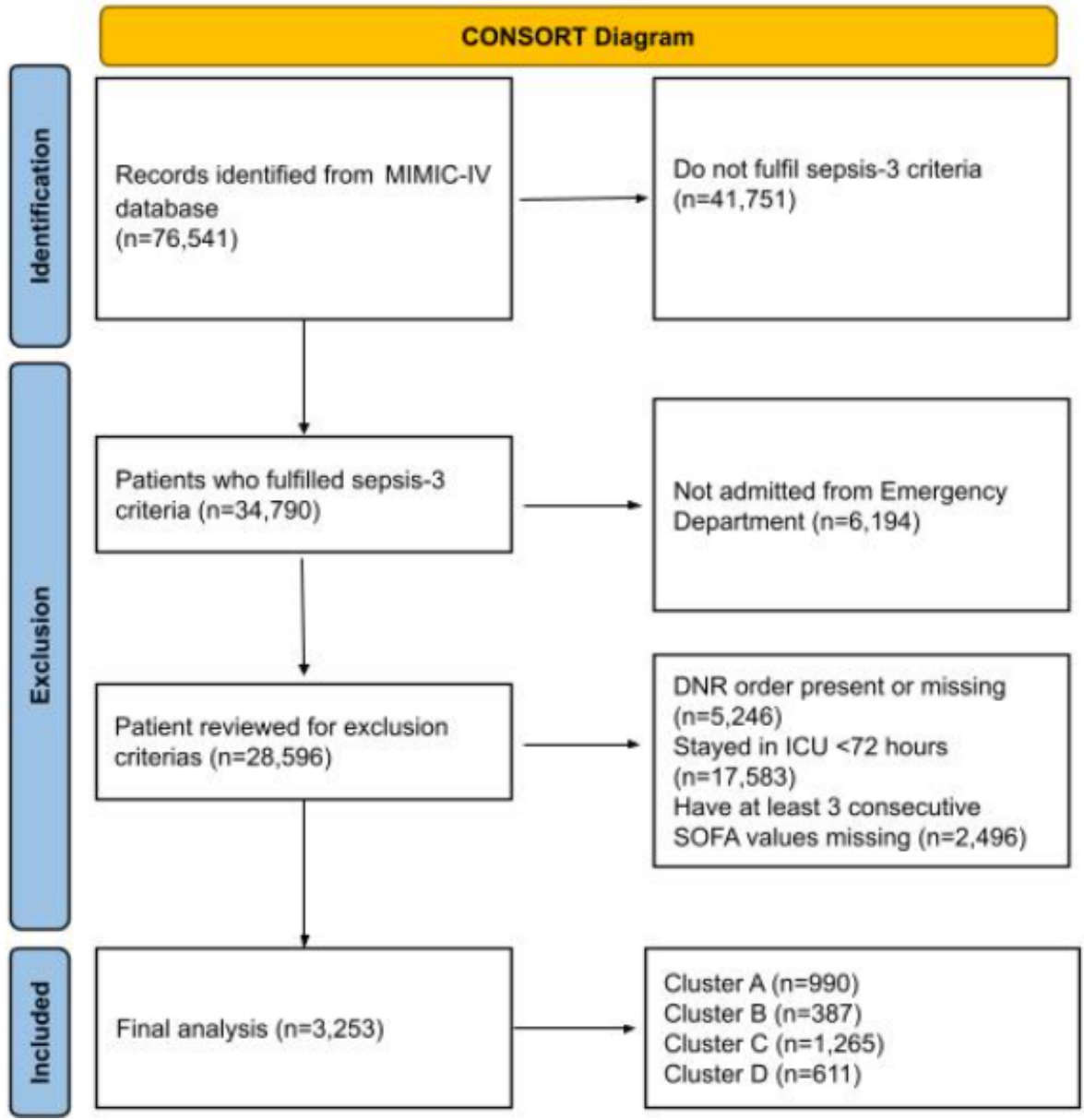

**Figure 1.** CONSORT diagram.

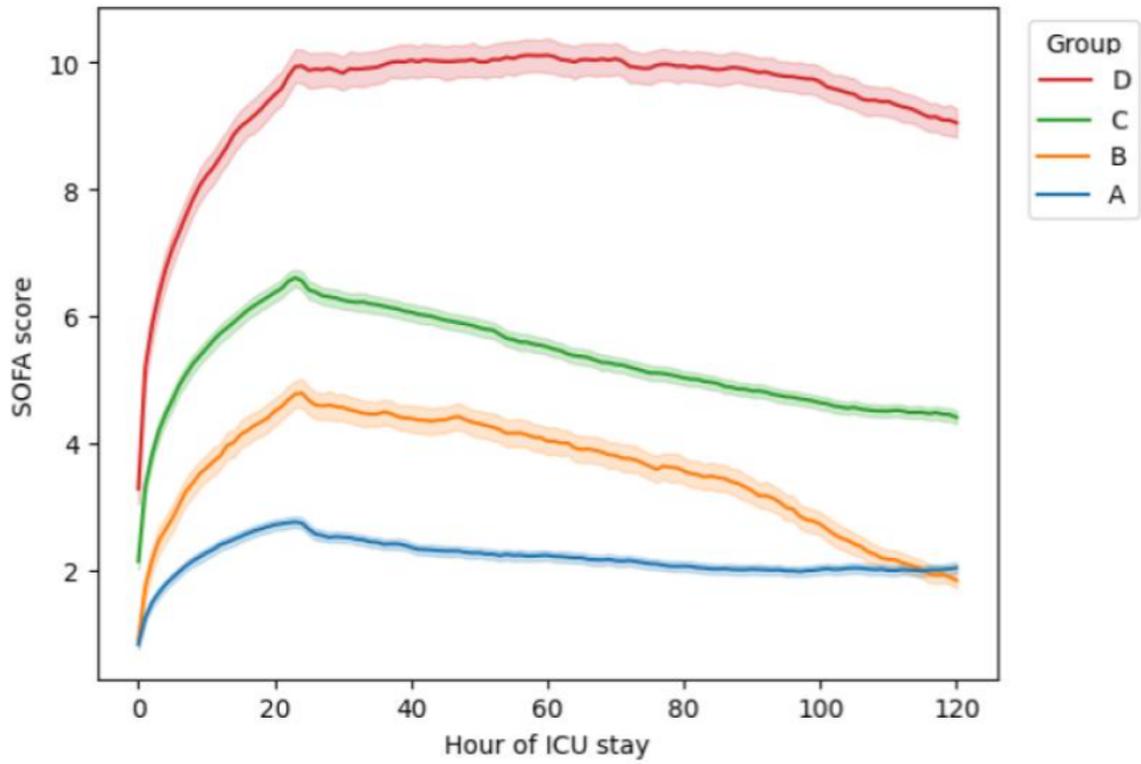

**Figure 2.** Four trajectories clusters of the first 72-hour time series SOFA score based on Group-based Trajectory Modelling (GBTM).

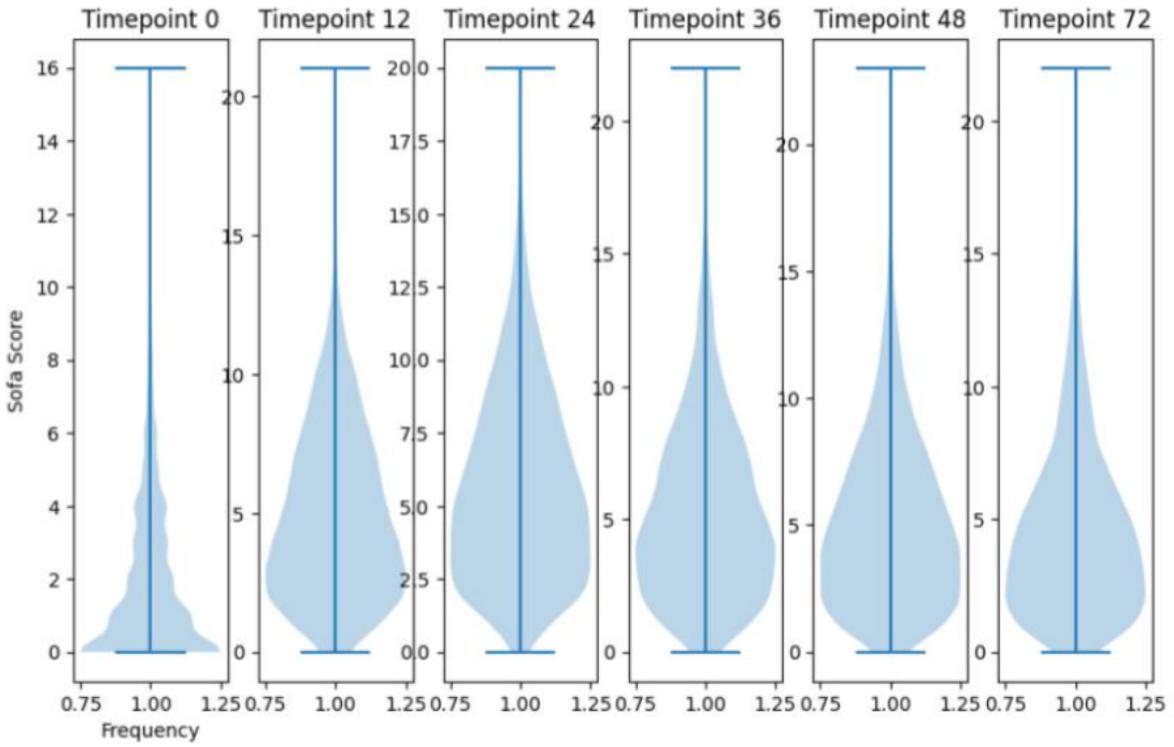

**Figure 3.** Violin plot of SOFA scores at set time points (hours).

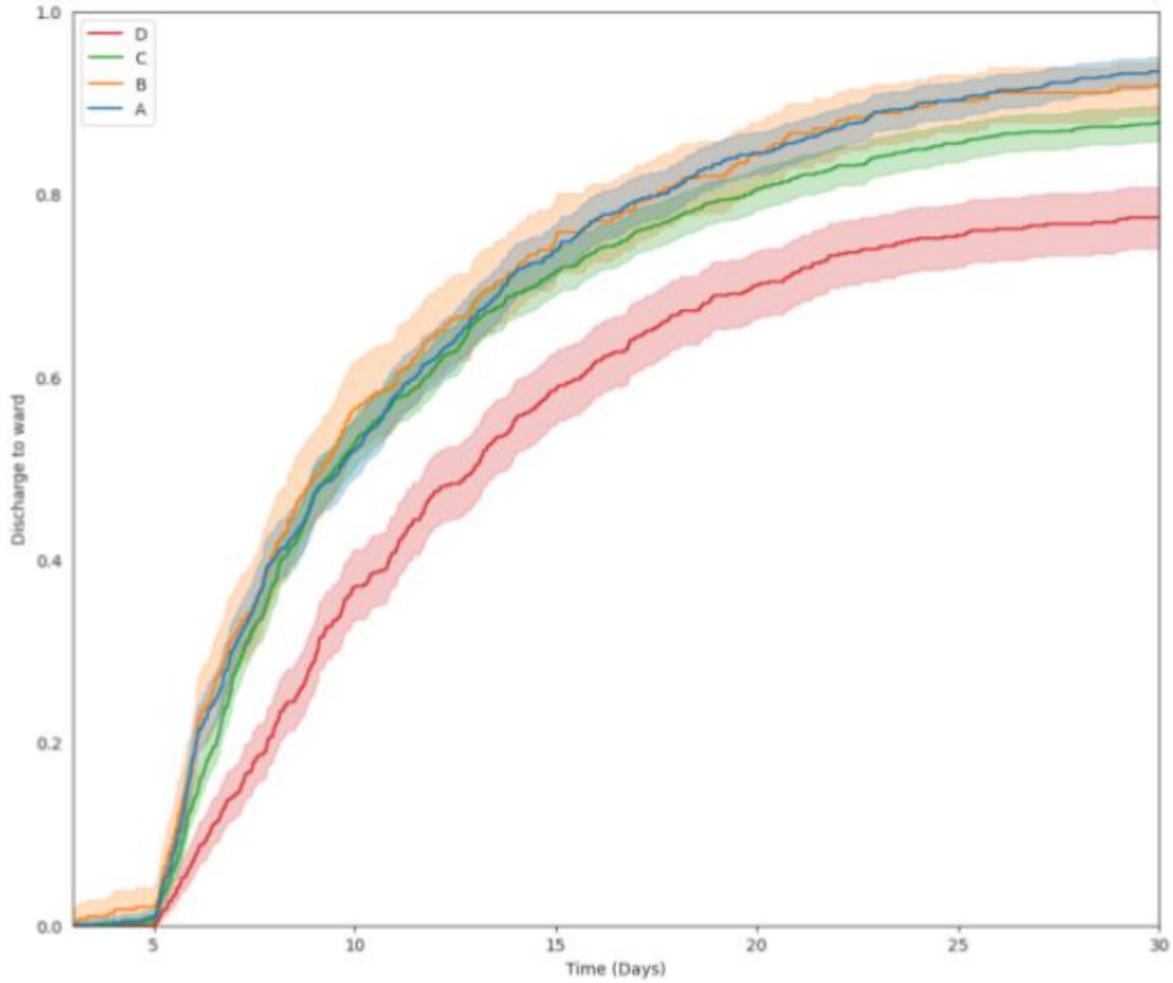

**Figure 4.** Discharge of patients from ICU to the ward stratified by trajectory clusters (p=0.701).

**Table 1:** Patient characteristics stratified by different cluster groups.

|  | A (n=990) | B (n=387) | C (n=1,265) | D (n=611) | p-value |
|---|---|---|---|---|---|
| Admission Age | 63 (50.0-75.0) | 63 (49.0-73.0) | 65 (53.0-76.0) | 62 (50.0-72.0) | <0.001 |
| BMI (kg/m$^2$) | 26.7 (23.3-31.6) | 28.0 (24.4-33.1) | 28.0 (24.0-33.4) | 28.5 (24.6-33.5) | <0.001 |
| CCI | 5 ±2 | 5 ±2 | 6 ±2 | 6 ±2 | <0.001 |
| Gender (Female) | 461 (46.6%) | 157 (40.6%) | 523 (41.3%) | 228 (37.3%) | 0.002 |
| Ethnicity | | | | | 0.904 |
|     Blacks | 113 (11.4%) | 43 (11.1%) | 159 (12.6%) | 76 (12.4%) | |
|     Hispanics | 34 (3.4%) | 23 (5.9%) | 48 (3.8%) | 30 (4.9%) | |
|     Asians | 31 (3.1%) | 9 (2.3%) | 32 (2.5%) | 21 (3.4%) | |
|     White | 587 (59.3%) | 229 (59.2%) | 754 (59.6%) | 350 (57.3%) | |
|     Others/Unknown | 225 (22.7%) | 83 (21.4%) | 272 (21.5%) | 134 (21.9%) | |
| Insurance | | | | | 0.001 |
|     Medicaid | 93 (9.4%) | 43 (11.1%) | 102 (8.1%) | 59 (9.7%) | |
|     Medicare | 384 (38.8%) | 148 (38.2%) | 598 (47.3%) | 250 (40.9%) | |
|     Others | 513 (51.8%) | 196 (50.6%) | 565 (44.7%) | 302 (49.4%) | |
| Platelets *10$^9$/L | 221.2 ± 103.1 | 200.1 ± 98.2 | 179.1 ± 104 | 128.5 ± 88.9 | <0.001 |
| Total White Cells *10$^9$/L | 14.1 ± 6.7 | 15.3 ± 7.6 | 17 ± 17.5 | 18 ± 11.3 | <0.001 |
| Bicarbonate (mEq/L) | 22.5 ± 4.7 | 21.2 ± 4.7 | 19.7 ± 5.5 | 16.6 ± 5.5 | <0.001 |
| Creatinine (mg/dL) | 1 ± 0.5 | 1.2 ± 0.7 | 2 ± 2.1 | 3.2 ± 4.1 | <0.001 |
| Bilirubin (mg/dL) | 0.7 ± 0.6 | 0.8 ± 0.6 | 1.1 ± 1.8 | 3.2 ± 5.7 | <0.001 |

BMI; Body Mass Index, CCI; Charlson Comorbidity Index, LOS; Length of Stay, ICU; Intensive care unit.

Median (IQR); Mean±SD; Number (%)

Kruskal-Wallis One-Way ANOVA

**Table 2.** Outcomes stratified by four trajectory clusters.

| | A (n=990) | B (n=387) | C (n=1,265) | D (n=611) | p-value |
|---|---|---|---|---|---|
| LOS ICU (days) | 9.2 (6.5-14.2) | 8.9 (6.3-13.8) | 9.0 (6.8-14.0) | 11.0 (7.8-16.1) | <0.001 |
| LOS Hospital (days) | 15.6 (10.7-22.8) | 16.0 (10.7-23.1) | 15.6 (11.0-22.3) | 18.6 (13.0-26.7) | <0.001 |
| ICU mortality | 28 (2.8%) | 11 (2.8%) | 89 (7.0%) | 96 (15.7%) | <0.001 |
| Hospital mortality | 55 (5.6%) | 18 (4.7%) | 138 (10.9%) | 122 (20.0%) | <0.001 |
| ICU readmissions | 104 (10.5%) | 55 (14.2%) | 142 (11.2%) | 75 (12.3%) | 0.242 |

| Discharge to ward by 1 week | 305 (30.8%) | 124 (32%) | 351 (27.7%) | 88 (14.4%) | <0.001 |
| Discharge to ward by 2 weeks | 717 (72.4%) | 284 (73.4%) | 888 (70.2%) | 341 (55.8%) | <0.001 |

LOS; Length of Stay, ICU; Intensive care unit.

Median (IQR); Number (%)

Kruskal-Wallis One-Way ANOVA